\documentclass[]{spie}  

\usepackage{amsmath,amsfonts,amssymb}
\usepackage{graphicx}
\usepackage[colorlinks=true, allcolors=blue]{hyperref}

\title{Modeling coronagraphic extreme wavefront control systems for high contrast imaging in ground and space telescope missions}

\author[a,b]{Jennifer Lumbres}
\author[a]{Jared Males}
\author[c]{Ewan Douglas}
\author[a]{Laird Close}
\author[a,b,d,e]{Olivier Guyon}
\author[c]{Kerri Cahoy}
\author[c]{Ashley Carlton}
\author[c]{Jim Clark}
\author[f]{David Doelman}
\author[g]{Lee Feinberg}
\author[a,b]{Justin Knight}
\author[c]{Weston Marlow}
\author[a,b]{Kelsey Miller}
\author[a]{Katie Morzinski}
\author[f]{Emiel Por}
\author[a,b]{Alexander Rodack}
\author[a,b]{Lauren Schatz}
\author[f]{Frans Snik}
\author[b]{Kyle Van Gorkom}
\author[f]{Michael Wilby}
\affil[a]{Steward Observatory, The University of Arizona, Tucson, AZ, USA}
\affil[b]{College of Optical Sciences, The University of Arizona, Tucson, AZ, USA}
\affil[c]{Department of Aeronautics and Astronautics, Massachusetts Institute of Technology, Cambridge, MA, USA}
\affil[d]{National Institutes of Natural Sciences, Subaru Telescope, National Observatory of Japan, Hilo, HI, USA}
\affil[e]{National Institutes of Natural Sciences, Astrobiology Center, Mitaka, Japan}
\affil[f]{Leiden Observatory, Leiden University, Leiden, Netherlands}
\affil[g]{NASA Goddard Space Flight Center, Greenbelt, MD, USA}

\authorinfo{Further author information:\\J.L. E-mail: jlumbres@optics.arizona.edu}

\begin{document} 
\maketitle

\begin{abstract}
The challenges of high contrast imaging (HCI) for detecting exoplanets for both ground and space applications can be met with extreme adaptive optics (ExAO), a high-order adaptive optics system that performs wavefront sensing (WFS) and correction at high speed. We describe two ExAO optical system designs, one each for ground-based telescopes and space-based missions, and examine them using the angular spectrum Fresnel propagation module within the Physical Optics Propagation in Python (POPPY) package. We present an end-to-end (E2E) simulation of the MagAO-X instrument, an ExAO system capable of delivering 6$\times10^{-5}$ visible-light raw contrast for static, noncommon path aberrations without atmosphere. We present a laser guidestar (LGS) companion spacecraft testbed demonstration, which uses a remote beacon to increase the signal available for WFS and control of the primary aperture segments of a future large space telescope, providing on order of a factor of ten factor improvement for relaxing observatory stability requirements. The LGS E2E simulation provides an easily adjustable model to explore parameters, limits, and trade-offs on testbed design and characterization.
\end{abstract}

\keywords{adaptive optics, wavefront control, Fresnel propagation, testbed modeling}

\section{INTRODUCTION}
\label{sec:intro}  


Large segmented aperture ground and space based telescopes are undergoing development to enable direct imaging for extrasolar planets. For ground-based systems, the Giant Magellan Telescope (GMT) is planned for first light in the mid-2020's. Future space-based segmented aperture telescope missions, such as LUVOIR (Large UV Optical Infrared Surveyor), are expected to have picometer-level observatory stability requirements for detecting Earth-like planets within the habitable zone\cite{lgs_Ruane}. The challenges of high contrast imaging for both ground and space applications can be met with extreme adaptive optics (ExAO), a high-order (${\geq}$ 1000 degrees of freedom) adaptive optics (AO) system which performs wavefront sensing and correction at high speed (${\geq}$ 1 kHz). 

We describe two extreme wavefront control systems, one for ground and another for space, currently in development. These optical system designs' capabilities for exoplanet direct detection are examined using the angular spectrum Fresnel propagation module within the Physical Optics Propagation in PYthon (POPPY) package. We present an E2E simulation of the MagAO-X instrument\cite{magaox_jared2018}, an NSF-funded ExAO upgrade for the Magellan Clay 6.5-meter telescope at Las Campanas Observatory in Chile. The MagAO-X simulation implements simulated optical surfaces generated from power spectral densities of each optical element and a vector Apodizing Phase Plate (vAPP) coronagraph\cite{vapp_frans_2012, vapp_otten, vapp_frans_spie}. MagAO-X's contrast is measured in the vAPP dark hole region after using a deformable mirror (DM) correction. For large aperture segmented space telescopes, we analyze a laser guidestar\cite{lgs_weston,lgs_spie2018} (LGS) companion spacecraft which uses a remote beacon to increase the signal available for wavefront sensing and control of the primary aperture segments. The LGS (potentially as small as a 6U or 27U cubesat) will fly in formation at a distance from the telescope (such as 10,000 km to 80,000 km), with the goal of increasing the speed and performance of the wavefront control system and relaxing the stringent stability requirements of a future large space telescope\cite{lgs_ewan_inprep}. The LGS demonstration testbed is currently being built at the Extreme Wavefront Control Laboratory (EWCL) at University of Arizona. The E2E simulation of the LGS testbed is to have an easily adjustable model to explore parameters, limits, and trade-offs on testbed design and characterization.  With MagAO-X and LGS, we show how extreme wavefront control efforts on both ground and space systems will enable high-contrast exoplanet direct imaging.

\section{SOFTWARE: PHYSICAL OPTICS PROPAGATION IN PYTHON (POPPY)}
\label{sec:poppy}
The software used for Fresnel analysis is POPPY \cite{poppy_SPIE2018, poppy_jwst} . The POPPY source code is available online for free\footnote{http://www.github.com/mperrin/poppy}. The POPPY framework allows users to build an optical system composed of multiple planes (pupils, images) from a library of optical element classes. POPPY can model both Fraunhofer and Fresnel diffraction for wavefront propagation through an optical system. Unlike ray tracing software, POPPY uses the paraxial approximation and assumes perfectly focusing optics. 


\subsection{Optical Elements}
\label{subsec:poppy_elements}
Since POPPY assumes that all optics inserted in the system will focus perfectly, many otherwise complex optical elements are represented in simple form. Parabolic mirrors are treated as quadratic lenses with a focal length parameter. Detection planes (focal planes, cameras), flat mirrors, and deformable mirrors are represented as scalar transmission locations.

\begin{figure} [ht]
   \begin{center}
   \begin{tabular}{c} 
   \includegraphics[height=5cm]{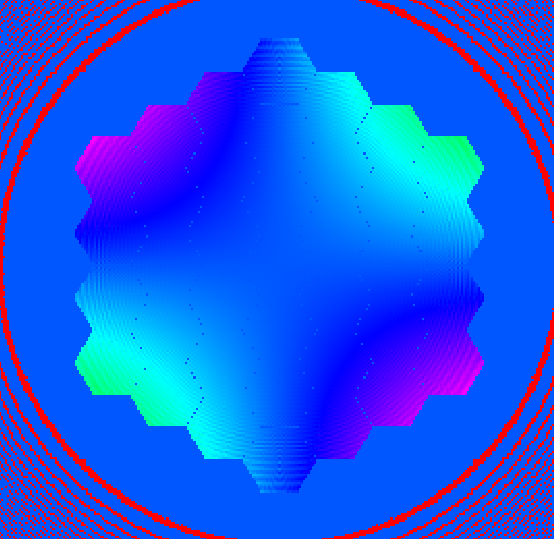}
	\end{tabular}
	\end{center}
   \caption[example] 
   { \label{fig:poppy_segDM}
Sample custom optical element built using POPPY built-in modules. This is a segmented mirror with astigmatism induced, which is used for LGS in Sec.~\ref{sec:lgs}.}
   \end{figure}

POPPY has built-in functions that can build custom optical elements as transmissive or OPD phase surfaces. It can also induce an aberration as a Zernike wavefront error (ZWFE). Fig.~\ref{fig:poppy_segDM} features a segmented mirror generated with astigmatism induced on the surface. More details on this are featured in the LGS simulation (Sec.~\ref{sec:lgs}).

\begin{figure} [ht]
   \begin{center}
   \begin{tabular}{c} 
   \includegraphics[height=5cm]{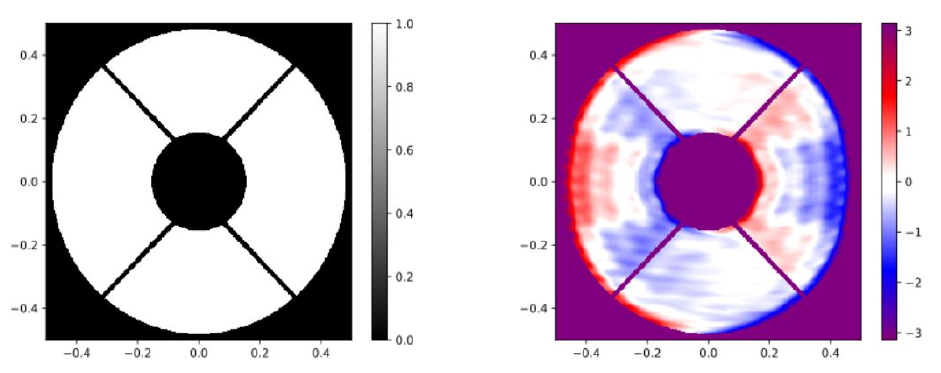}
	\end{tabular}
	\end{center}
   \caption[example] 
   { \label{fig:poppy_vAPP}
The vAPP coronagraph mask to be used in MagAO-X. (Left) vAPP amplitude component, which is passed in POPPY as a transmission mask; (Right) vAPP phase map (units: radians), which is converted to OPD and passed into POPPY as an OPD surface map. The transmission and OPD surface maps are superposed together in POPPY to form a complex phase mask.}
   \end{figure}

POPPY also allows the user to insert custom amplitude transmissive or optical path difference (OPD) elements into the system, thereby letting the user induce aberration at their discretion. The surfaces are applied when the optical element is declared in POPPY. Fig.~\ref{fig:poppy_vAPP} features the vAPP coronagraph, which utilizes POPPY's custom amplitude transmission and OPD capabilities (see Sec.~\ref{subsec:magaox_model} for more details on the surfaces used in MagAO-X). 

\section{GROUND: MAGELLAN EXTREME ADAPTIVE OPTICS (MagAO-X)}
\label{sec:magaox}

The goal of the MagAO-X Fresnel propagation simulation is to characterize the MagAO-X instrument layout's influence on the dark hole (DH) contrast levels produced by the vAPP coronagraph. One major contributor to the contrast level is the optical surface imperfections of each optic. There are 23 optical element surfaces evaluated in the coronagraphic E2E path, excluding the woofer-tweeter DMs and vAPP coronagraph. From here, we break the analysis into 2 major parts: determining the lowest possible contrast level for an E2E simulation (see section \ref{subsec:magaox_dh}) and verifying the optical specification requirements set (see section \ref{subsec:magaox_verify}).
 
\subsection{MagAO-X Instrument Model Description}
\label{subsec:magaox_model}
\begin{figure} [ht]
   \begin{center}
   \begin{tabular}{c} 
   \includegraphics[height=6.5cm]{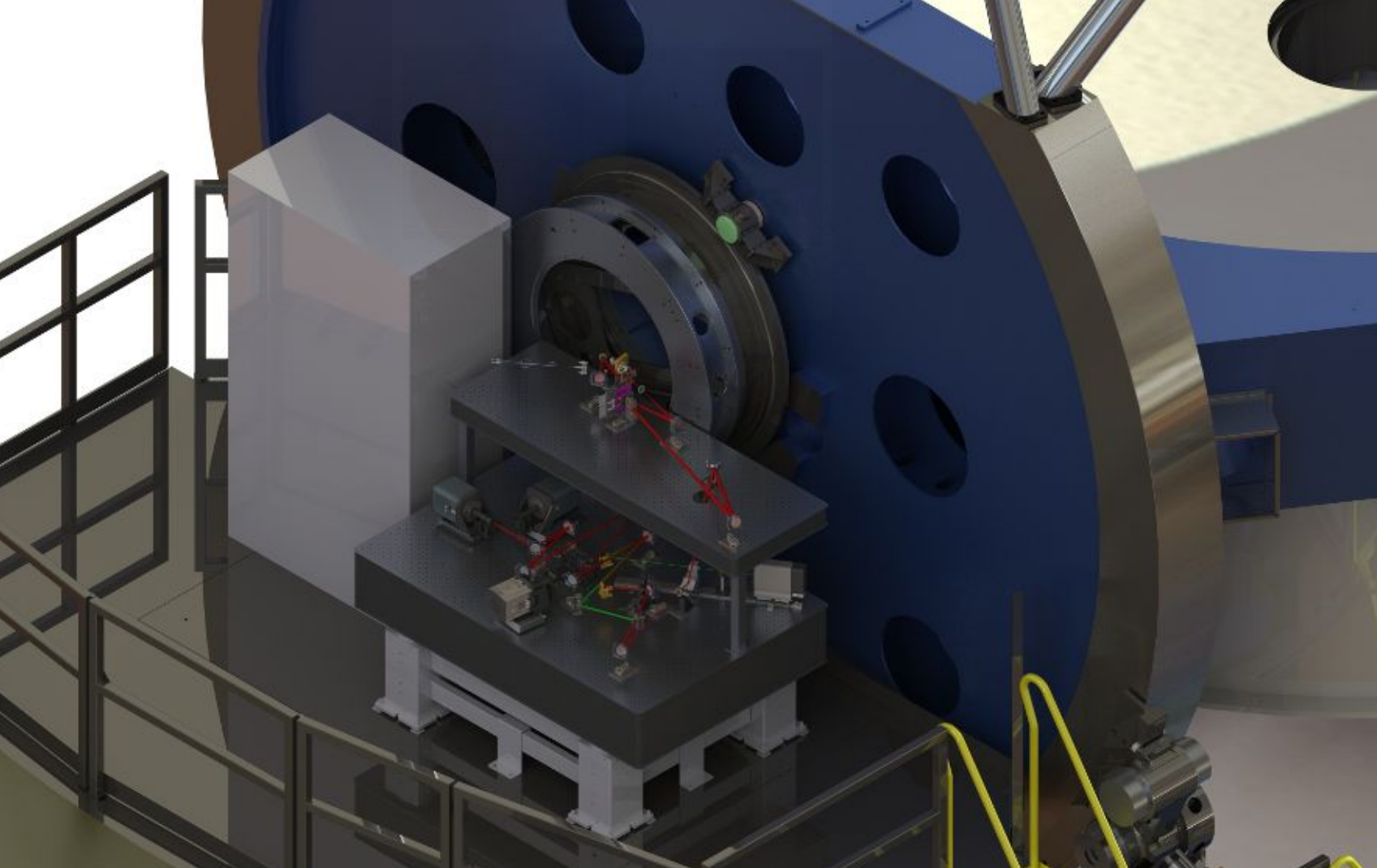}
	\end{tabular}
	\end{center}
   \caption[example] 
   { \label{fig:magaox_render}
A rendering of the MagAO-X instrument on the Magellan Clay telescope platform. The MagAO-X instrument is a two-layer bench, shown here with the dust covers removed. Fig.~\ref{fig:magaox_raytrace} shows the raytrace diagram of the optical elements featured on each bench level.}
   \end{figure}

\begin{figure} [ht]
   \begin{center}
   \begin{tabular}{c} 
   \includegraphics[height=6.5cm]{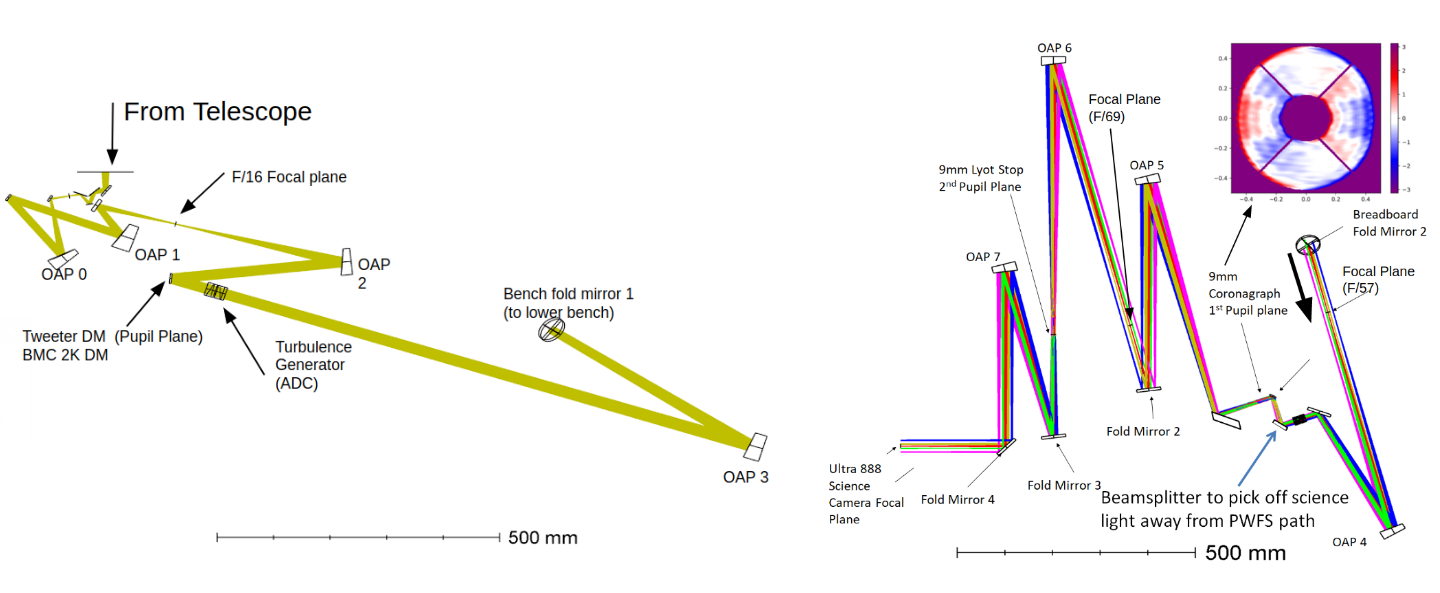}
	\end{tabular}
	\end{center}
   \caption[example] 
   { \label{fig:magaox_raytrace}
ZEMAX raytrace of MagAO-X instrument. (Left) Top bench ray trace, which begins at ``From Telescope''. (Right) Bottom bench raytrace, which begins at ``Breadboard Fold Mirror 2'' with a large arrow pointing at direction of raytrace.}
   \end{figure}

In MagAO-X, the Fresnel propagation model uses the ZEMAX optical raytrace prescriptions set as the propagation distances between optics \cite{Laird_magaox}. The E2E propagation path used is through the coronagraph science path, so the pyramid WFS\cite{lauren_pywfs} is not included in the Fresnel analysis. See Fig.~\ref{fig:magaox_render} for the MagAO-X instrument rendering. Fig.~\ref{fig:magaox_raytrace} features the ZEMAX raytrace analysis of the MagAO-X instrument used for the Fresnel propagation analysis presented here.

\begin{figure} [ht]
   \begin{center}
   \begin{tabular}{c} 
   \includegraphics[height=7cm]{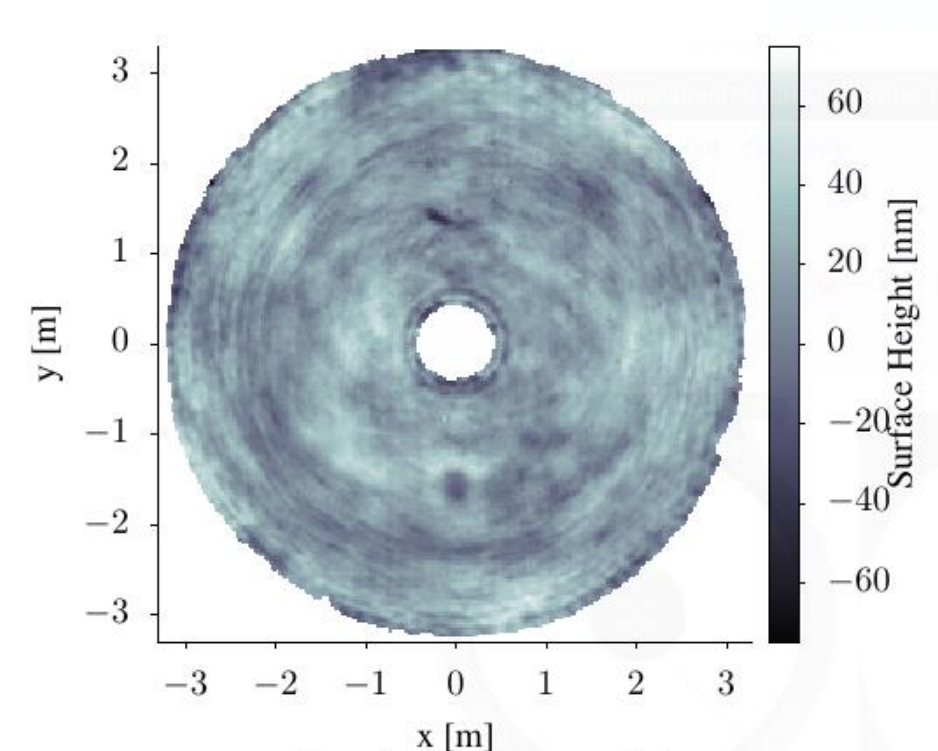}
	\end{tabular}
	\end{center}
   \caption[example] 
   { \label{fig:magClay_PM}
Measured surface map of Magellan Clay primary mirror}
   \end{figure}

\begin{figure} [!ht]
   \begin{center}
   \begin{tabular}{c} 
   \includegraphics[height=5cm]{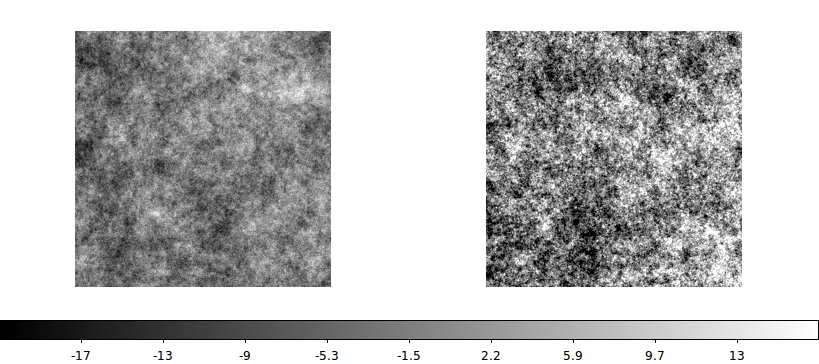}
	\end{tabular}
	\end{center}
   \caption[example] 
   { \label{fig:PSD_samp}
Sample PSD surface maps generated for MagAO-X in OPD units (nm). (Left) Flat mirror (F-1); (Right) OAP mirror (O-0)}
   \end{figure}

Surface maps are generated using the MagAO-X optics specifications. The actual surface map of the Magellan Clay telescope primary mirror (M1) is used (see Fig.~\ref{fig:magClay_PM}). For the other optics, PSDs are generated with the appropriate parameters and normalization. M2 and M3 are based on the known as-built surface specifications. For the new optics we used ${\lambda}$/50 reflected wavefront error for the OAPs and ${\lambda}$/100 (PV) surfaces for the flat mirrors. The PSDs were used to generate surface maps by using standard Fourier convolution with Gaussian white noise. See Fig.~\ref{fig:PSD_samp} for sample surface maps.

The vAPP coronagraph used for high contrast characterization is for single DH generation. The initial DM surfaces were simulated as perfect surfaces, as the DM surface maps\cite{kyle_DM} were not available at the time. All of these tests are performed assuming a bright source with no aberration. The wavelength used for the MagAO-X Fresnel analysis is H$\alpha$.

\subsection{Dark Hole Contrast Characterization}
\label{subsec:magaox_dh}
There are 3 different DH contrast characterizations calculated: the optimal DH contrast, the initial DH contrast for an open loop E2E setup with optical surfaces, and the DH contrast for a single-iteration closed loop E2E setup with optical surfaces applied. The region of interest (ROI) used for calculating DH contrast in each test is a 54 x 54 pixel box (approximately 3 ${\lambda}$/D to 9 ${\lambda}$/D) inscribed within the DH region. The DH contrast is calculated by averaging the individual pixels inside the ROI box of the fully-normalized image. The boxed ROI design choice was selected for quick comparison analysis. Future work will cover the entire dark hole. Each of the procedures are described and the contrasts values compared.

\subsubsection{Optimal Dark Hole Contrast}
\label{subsubsec:magaox_dh_test1}
\begin{figure} [ht]
   \begin{center}
   \begin{tabular}{c} 
   \includegraphics[height=5.5cm]{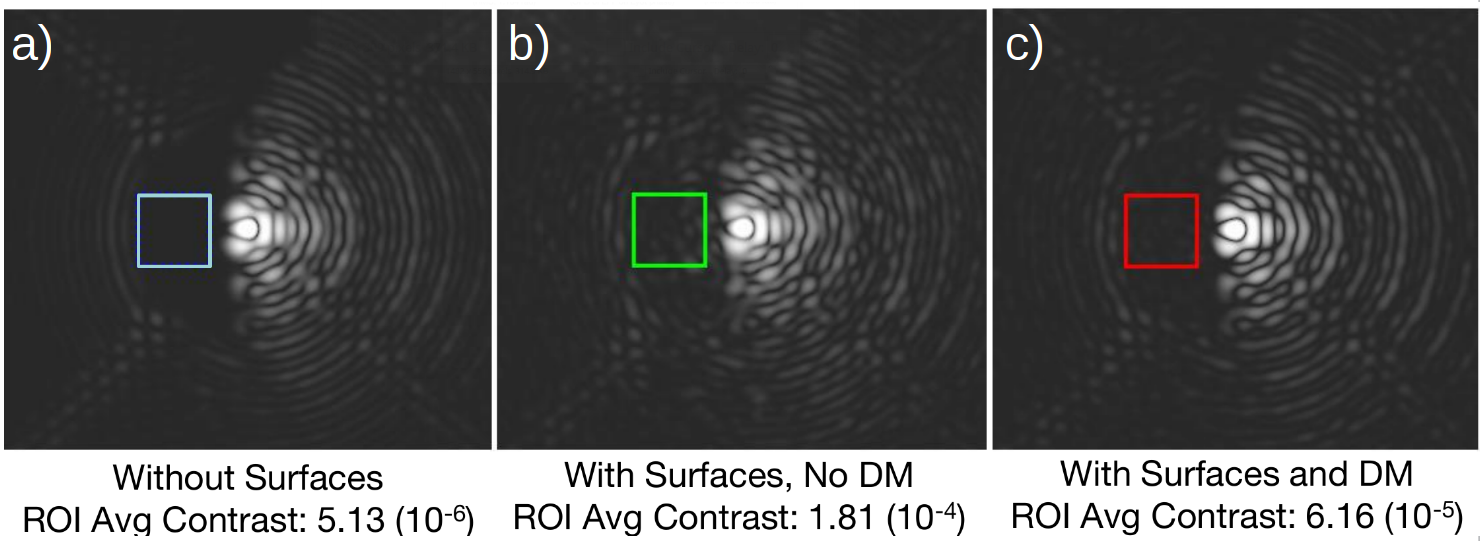}
	\end{tabular}
	\end{center}
   \caption[example] 
   { \label{fig:magaox_dhcontrast}
MagAO-X coronagraphic PSFs generated through Fresnel propagation. The colored box represents the region of interest used in calculating the DH contrast for each particular test.}
   \end{figure}

The optimal DH contrast is calculated by running an open loop E2E simulation of the instrument but with perfect (unaberrated) surfaces set on each optical element. An unaberrated surface is set in POPPY by not applying a surface map at each optic, which is labeled in this paper as the ``without surfaces'' DH contrast. This contrast level sets the MagAO-X contrast limit and serves as a reference for the other tests. Fig.~\ref{fig:magaox_dhcontrast}a features the coronagraphic PSF for the ``without surfaces'' test, which was found to be 6.16$\times10^{-6}$ and the contrast limit of MagAO-X using the vAPP coronagraph.

\subsubsection{Open Loop E2E With Surfaces Applied}
\label{subsubsec:magaox_dh_test2} 
The second test is the initial DH contrast calculated for an open loop E2E simulation with all available optical surfaces inserted. Since it is open loop, the DMs remain as perfect surfaces and do not apply a correction. Its label is ``With Surfaces, No DM''. This was done to observe how the planned optical surfaces affect the DH contrast and whether their specifications are sufficient enough to form a high contrast DH. Fig.~\ref{fig:magaox_dhcontrast}b features the ``With Surfaces, No DM'' coronagraphic PSF, which was found to be 1.81$\times10^{-4}$ contrast, which is 2 magnitudes worse than the optimal DH contrast.

\subsubsection{Single Iteration Closed Loop Correction E2E With Surfaces Applied}
\label{subsubsec:magaox_dh_test3}
The third test is another E2E simulation with optical surfaces implemented, but includes a single iteration closed-loop correction at the Tweeter DM surface. Its label is ``With Surfaces and DM''. The goal of this test is to check if the DM can handle correcting the surface imperfections without losing too much stroke as well as remedy the 2 magnitude contrast degredation from the open loop E2E test.

\begin{figure} [!ht]
   \begin{center}
   \begin{tabular}{c} 
   \includegraphics[height=5.5cm]{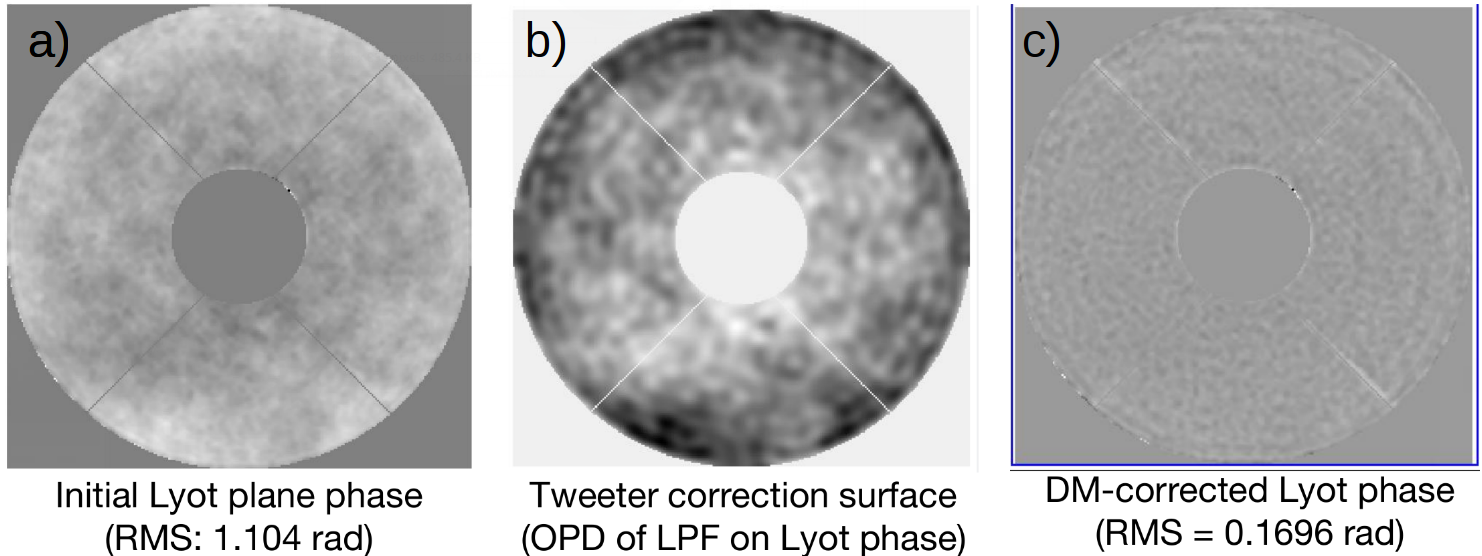}
	\end{tabular}
	\end{center}
   \caption[example] 
   { \label{fig:magaox_DMcorr}
Phase and OPD maps used in DM correction. a) Phase seen at Lyot plane; b) OPD correction derived by WFS and applied to BMC-2K ``tweeter'' DM; c) Residual wavefront error after DM correction applied.  }
   \end{figure}

The pick-off location for the DM correction is at the Lyot stop because it is the last pupil plane before the coronagraphic PSF and detects most of the surface errors in the optics stream. POPPY calculates and returns the complex field at each propagation plane, so the WFS signal is the phase generated at the Lyot stop (see Fig.~\ref{fig:magaox_DMcorr}a). The phase data is then moved to the Fourier domain, where the spatial frequencies beyond 12 ${\lambda}$/D are filtered out from the Boston Micromachines Corporation 2000 actuator (BMC-2K) ``tweeter'' DM. The filtered data is returned to the phase domain where it is converted to OPD and inverted before being inserted as the correction surface map for the Tweeter DM (see Fig.~\ref{fig:magaox_DMcorr}b). The whole system is run again from the start with the correction DM map in place. A single iteration of this correction reduced the Lyot plane phase RMS error from 1.104 rad to 0.1696 rad (see Fig.~\ref{fig:magaox_DMcorr}c). The DH contrast for the single iteration closed loop correction improves from 1.81$\times10^{-4}$ to 6.16$\times10^{-5}$ (see Fig.~\ref{fig:magaox_dhcontrast}c), which is a whole magnitude of improvement from the open loop E2E analysis with surfaces. This contrast is considered as the MagAO-X DH contrast characterization with the vAPP coronagraph.

\subsection{Verification of Optical Element Specifications}
\label{subsec:magaox_verify}
A second analysis performed was to verify the optical element specifications for the MagAO-X instrument. While the DH contrast characterization tests performed in Sec.~\ref{subsec:magaox_dh} showed that the optical element specifications can sufficiently produce a DH, the next goal is examining if there are any individual optical element(s) that affect the DH contrast. We perform this task to ensure that the optical elements procured will be arranged in an optimal manner for the MagAO-X instrument.

For this test, the analyzed powered optical element has its surface aberrations removed to simulate an ideal, unaberrated optical surface while while all the other optical elements' surfaces remain the same. Replacing the tested optical element surface will show the diffraction effect produced by the optical element's surface onto the DH contrast. A DH contrast is generated for the removed optical element surface and repeated across each of the 23 optical elements. Additionally, this process is repeated using 5 sets of surface maps generated from the same PSD set, where the DH contrast is averaged across each set per tested optical element surface. 

\begin{figure} [!ht]
   \begin{center}
   \begin{tabular}{c} 
   \includegraphics[height=8.5cm]{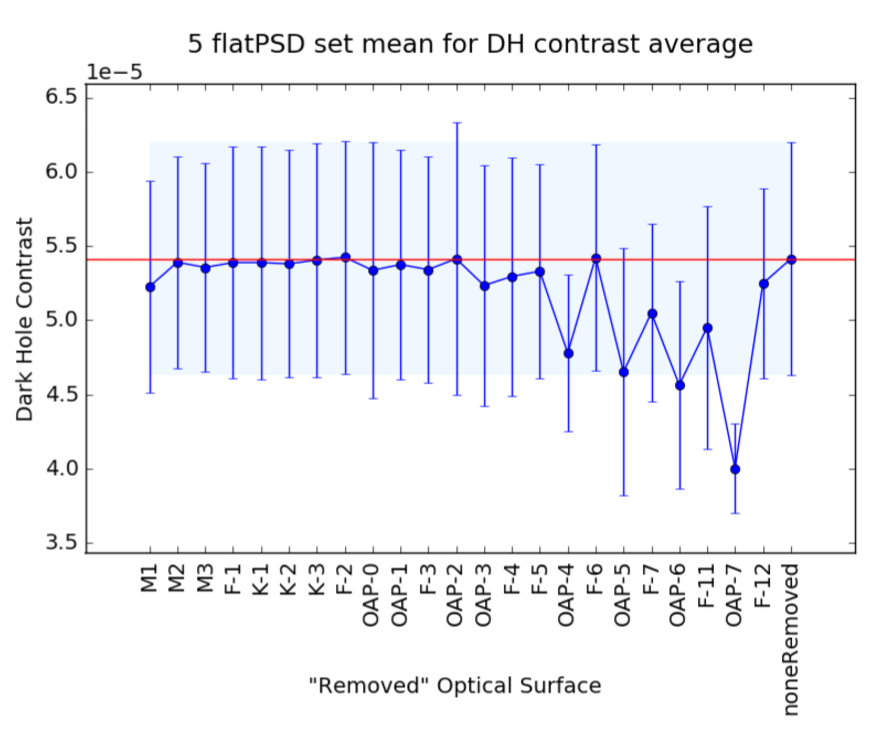}
	\end{tabular}
	\end{center}
   \caption[example] 
   { \label{fig:magaox_verif}
MagAO-X verification of optical element specifications test. The lower the DH contrast (vertical axis) result, the further degradation of the DH contrast due to mid-spatial frequency aberration content of each optical surface added into the system. The error bars of each tested optic are within one standard deviation of the contrast produced in 5 surface map sets.}
   \end{figure}

The results of the individual optical element surface test are featured in Fig.~\ref{fig:magaox_verif}. The red horizontal line is the baseline reference DH contrast averaged across the tests where no optical element surfaces were removed (therefore labeled ``noneRemoved''). The light blue box is the 1 standard deviation regime produced by the noneRemoved reference. Each tested optical element surface average DH contrast that lies inside the light blue box show that it performs on par with the noneRemoved reference. Therefore, the optical surface quality for that optical element is verified to meet performance.

However, it was found that when OAP7 is replaced with a perfect surface, the DH contrast performance improvement exceeds 1 standard deviation of the noneRemoved reference. This shows that the surface quality of OAP7 will have a critical impact on the MagAO-X system performance. The reason for this phenomenon is that OAP7 is ``non-common path'' (NCP) with the WFS because it is positioned after the pick-off point to the WFS. Since it is NCP with the WFS, the WFS is blind to it, therefore the DM cannot correct it. A simple solution to this problem is to insert the best surface quality OAP for OAP7 during MagAO-X instrument assembly, which is currently underway\cite{alexH_kmirror, maggie_spie}. Additionally, it can be compensated for by speckle nulling and/or another form of focal plane wavefront sensing, which is currently being analyzed for MagAO-X\cite{kelsey_spie}.

\section{SPACE: LASER GUIDE STAR FOR LARGE APERTURE SEGMENTED SPACE TELESCOPES}
\label{sec:lgs}


Our goal in conducting a Fresnel simulation of the LGS testbed is to have an easily adjustable model to explore parameters, limits, and trade-offs on testbed design and characterization. This in turn will affect the LGS mission parameters being designed by the Massachusetts Institute of Technology STAR Laboratory\cite{lgs_weston,lgs_ewan_inprep}. The initial model is described in ~\ref{subsec:lgs_model}. Sec.~\ref{subsec:lgs_fresnel} features the LGS correction procedure and preliminary results of the Fresnel propagation for the testbed design model.

\subsection{LGS Demonstration Testbed Model Description}
\label{subsec:lgs_model}
\begin{figure} [!ht]
   \begin{center}
   \begin{tabular}{c} 
   \includegraphics[height=6.5cm]{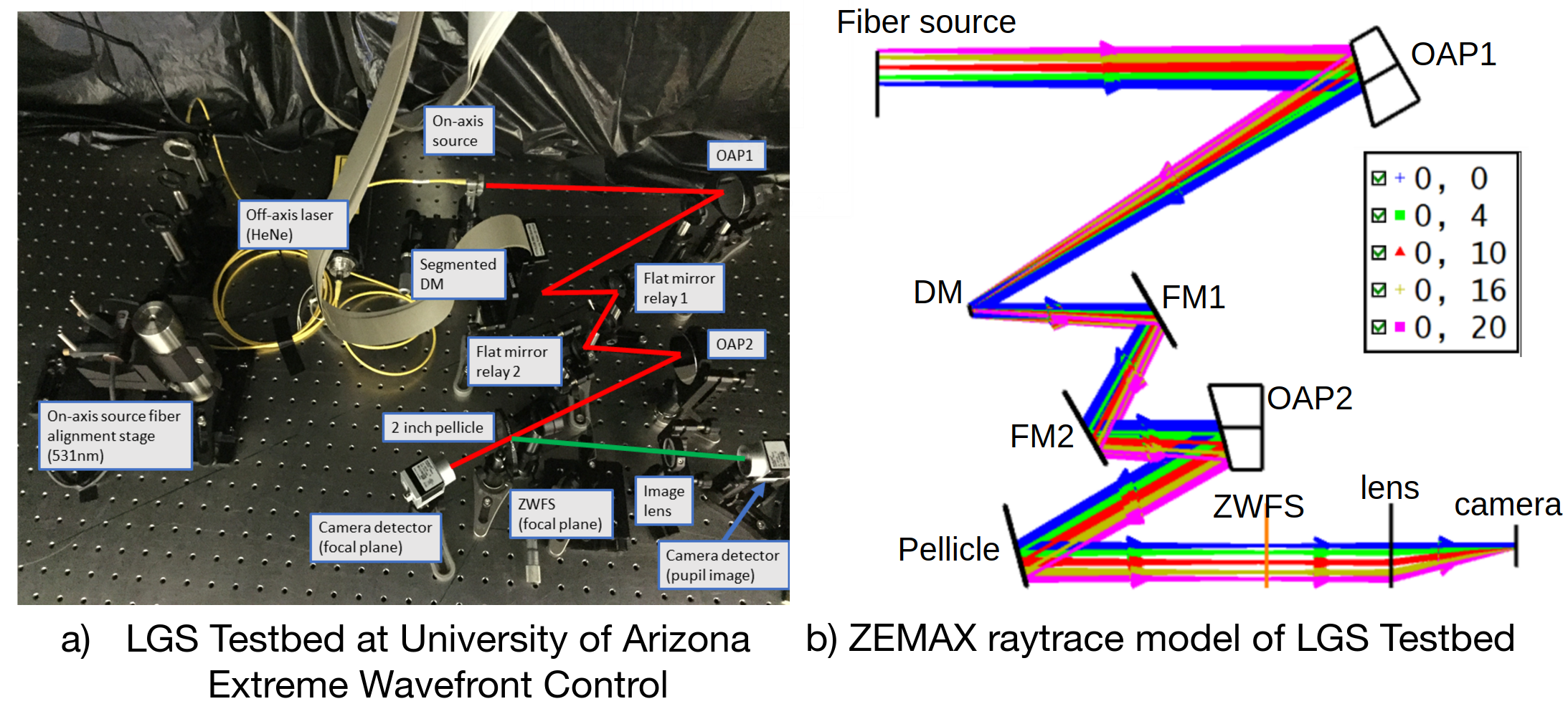}
	\end{tabular}
	\end{center}
   \caption[example] 
   { \label{fig:lgs_testbed}
LGS demonstration testbed (preliminary design) at University of Arizona EWCL. The off-axis source will be fed into the system using a beamsplitter cube in future designs.}
   \end{figure}

The LGS demonstration testbed is a single pupil relay system. Fig.~\ref{fig:lgs_testbed} shows the testbed at the University of Arizona EWCL and a ZEMAX optical design model. It has two fiber point sources to act as the on-axis target star and an off-axis LGS source. The wavelengths used are 531 nm for the target star and a HeNe (633 nm) for the LGS. We are using an IrisAO PTT-111L segmented DM at the first pupil plane to represent LUVOIR. There is a 7 mm inscribed aperture placed at the DM. A pellicle is placed before the first focal plane, where one path leads to a camera to image the on-axis source. The other is the WFS path, where a Zernike WFS\cite{zwfs_1942,zwfs_bloemhof} (ZWFS) is placed at the focal plane of the off-axis LGS rays. An imaging lens is placed between the ZWFS and camera to image the filtered segmented DM pupil.

The on-axis source is static and unaberrated, similar to MagAO-X. The off-axis source is modeled in POPPY using a ZWFE with tilt and defocus prescribed. The tilt value is based on the number of waves seen by LUVOIR for a 10 arc-seconds off-axis source. The defocus value is based on the LGS cubesat placed 50,000 km away from LUVOIR. 

Unlike the MagAO-X Fresnel propagation simulation which had optical element surface maps, the LGS simulation presented here does not include surface maps for the OAPs and flat mirrors. This will be implemented into future LGS Fresnel simulations in order to better calibrate testbed artifacts. The only surface map used in the LGS simulation is the generated aberrations for the DM. 

\subsection{Preliminary Fresnel Propagation Simulations of LGS Demonstration Testbed}
\label{subsec:lgs_fresnel}
\begin{figure} [ht]
   \begin{center}
   \begin{tabular}{c} 
   \includegraphics[height=7cm]{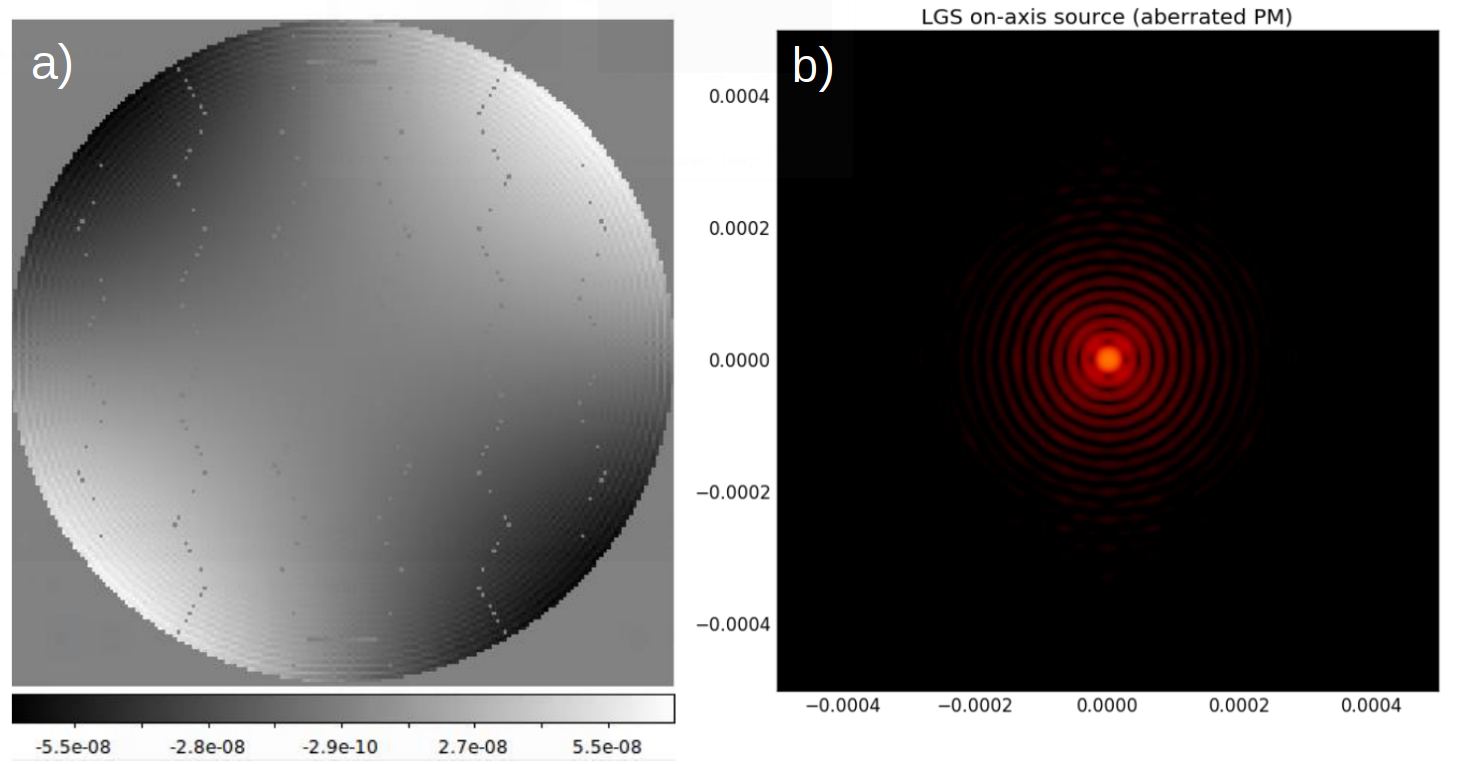}
	\end{tabular}
	\end{center}
   \caption[example] 
   { \label{fig:lgs_pm_aberr}
Simulated test setup for LGS correction. a) Aberrated segmented DM for on-axis source. This is masked by the 7mm inscribed aperture on the IrisAO PTT-111L. b) Simulated PSF produced by aberrated segmented DM. }
   \end{figure}

A single iteration closed-loop using the LGS source to correct the aberrated segmented primary mirror is described here. The on-axis aberrated segmented DM surface map is shown in Fig.~\ref{fig:lgs_pm_aberr}a and has 0.5 waves of astigmatism applied at 633nm.  Running the Fresnel propagation simulation produces the on-axis PSF in Fig.~\ref{fig:lgs_pm_aberr}b.

\begin{figure} [ht]
   \begin{center}
   \begin{tabular}{c} 
   \includegraphics[height=5cm]{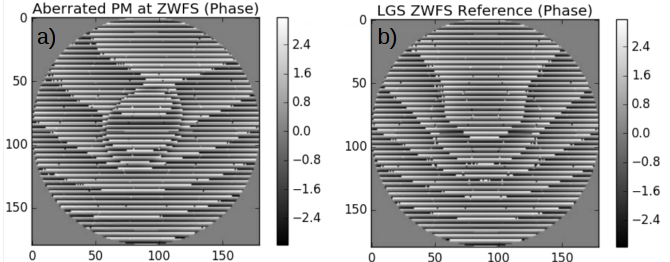}
	\end{tabular}
	\end{center}
   \caption[example] 
   { \label{fig:lgs_zwfs}
Phase measured by ZWFS: a) Aberration applied by DM; b) LGS phase calibration}
   \end{figure}

The LGS path follows from the ZWFE source and through the ZWFS route. The ZWFS image plane phase map seen by the LGS with the aberrated DM is shown in Fig.~\ref{fig:lgs_zwfs}a. To correct the on-axis aberrated segmented DM, the ZWFS image plane phase map needs to remove the phase map contributed by the LGS itself. The LGS calibration phase map is produced by propagating the LGS through the system assuming a flat segmented DM and is shown in Fig.~\ref{fig:lgs_zwfs}b.

To form a correction on the aberrated DM with the ZWFS phase, the LGS calibration phase map must be removed from the ZWFS image plane phase map, converted to OPD, inverted, and added to the aberrated DM surface map. This is currently undergoing analysis and will appear in future work. 



\section{CONCLUSION AND FUTURE WORK}

We show that through an E2E simulation of the MagAO-X instrument that MagAO-X is capable of delivering visible-light high contrast imaging with the vAPP coronagraph at 6.16$\times10^{-5}$ limited by static, non-common path aberrations at small angular separations (1-10 ${\lambda}$/D, 19-190 mas). The optical elements surfaces are verified to meet the specification requirements for DH generation. The MagAO-X optical elements have been ordered and have recently been received at University of Arizona Steward Observatory. They are currently undergoing testing with a Zygo Verifire Fizeau interferometer in the EWCL at University of Arizona. PSD profiles of these optical elements are currently underway for development and will be incorporated into the MagAO-X POPPY model. Once the PSD analysis has been completed, it will be incorporated as a module into POPPY.  Additionally, a dust analysis studying the effect on the dark hole contrast is currently being planned.

We demonstrated the E2E preliminary simulation results of the LGS concept study for correcting an aberrated segmented aperture. Future work with the LGS include multiple iterations of closed-loop correction, quantifying the improvement in order to demonstrate the feasibility of picometer level wavefront sensing, adding a coronagraph into the simulation, and exploring parameter space to determine the optimal WFS architecture. This will eventually converge for a more robust testbed model with the parameters and design trade-offs considered. Once the design has been finalized through the E2E simulation, it will be assembled and tested in the EWCL at University of Arizona. Laboratory results will be compared with the Fresnel simulation.

\acknowledgments 
 
We are grateful for the support we have received for MagAO-X and LGS. MagAO-X has been possible through NSF MRI research grant $\#$1625441 (\textit{Development of a Visible Wavelength Extreme Adaptive Optics Coronagraphic Imager for the 6.5 meter Magellan Telescope}, P.I.: Jared Males). LGS has been possible through NASA Early Stage Innovations grant $\#$NNX17AD07G (\textit{Laser Guide Star for Large Aperture Segmented Telescopes}, P.I.: Kerri Cahoy).

\bibliography{report} 
\bibliographystyle{spiebib} 

\end{document}